\documentclass[12pt,a4paper]{article}

 \usepackage{graphics}
 \usepackage{graphicx}
 \usepackage{afterpage}
 \usepackage{enumerate}

\begin{document}
 %%==========================================================

 \title{\bf Tropospheric ozone columns and ozone profiles \\ for Kiev in 2007}

  \author{
    A.V. Shavrina$^1$, Ya.V. Pavlenko$^1$,
    A.A. Veles$^1$,
    V.A. Sheminova$^1$,
    I.I. Synyavski$^1$,\\
    M.G. Sosonkin$^1$,
    Ya.O. Romanyuk$^1$,
    N.A. Eremenko$^1$,
    Yu.S. Ivanov$^1$,\\
    O.A. Monsar$^1$,
    M. Kroon$^2$}
\date{}
 \maketitle
 \begin{center}{
          $^1$Main Astronomical Observatory,
          National Academy of Sciences of the Ukraine,\\
           27 Akademika Zabolotnoho St.,
           03680 Kiev, Ukraine

           $^2$Royal Netherlands Meteorological
           Institute (KNMI), The Netherlands}
\end{center}

\begin{abstract}
{We report on ground-based FTIR observations being performed
within the framework of the ESA-NIVR-KNMI project 2907 entitled
``OMI validation by ground based remote sensing: ozone columns
and atmospheric profiles'' for the purpose of OMI data
validation. FTIR observations were performed during the time
frames August-October 2005, June-October 2006 and March-October
2007, mostly under cloud free and clear sky conditions and in
some days from early morning to sunset covering the full range
of solar zenith angles possible. Ozone column and ozone profile
data were obtained for the year 2005 using spectral modeling of
the ozone spectral band profile near 9.6 microns with the
MODTRAN3 band model based on the HITRAN-96 molecular absorption
database. The total ozone column values retrieved from FTIR
observations are biased low with respect to OMI-DOAS data by
8-10~DU on average, where they have a relatively small standard
error of about 2\%. FTIR observations for the year 2006 were
simulated by MODTRAN4 modeling. For the retrieval of ozone
column estimates and particularly ozone profiles from our FTIR
observations, we used the following data sources to as input
files to construct the (a priori ) information for the model:
satellite Aqua-AIRS water vapor and temperature profiles;
Aura-MLS stratospheric ozone profiles (version 1.5), TEMIS \cite{4}
climatological ozone profiles and the simultaneously performed
surface ozone measurements. Ozone total columns obtained from
our FTIR observations for year 2006 with MODTRAN4 modeling are
matching rather well with OMI-TOMS and OMI-DOAS data where
standard errors are 0.68\% and 1.11\%, respectively. AURA-MLS
data of version 2.2 which have become available in 2007 allow us
to retrieve tropospheric ozone profiles. For some days Aura-TES
tropospheric profiles were also available and were compared with
our retrieved profiles for validation. A preliminary analysis of
troposphere ozone variability was performed. Observation during
the time frame March-October demonstrate daily photochemical
variability of tropospheric ozone and reveal mixing processes
during the night.}
%\keywords{}
\end{abstract}

%%%%%%%%%%%%%%%%%%%%%%%%%%%%%%%%%%%%%%%%%%%%%%%%%%%%%%%%%%%%%%%%%%%%%%%%%%%%
\section{Introduction}
%%%%%%%%%%%%%%%%%%%%%%%%%%%%%%%%%%%%%%%%%%%%%%%%%%%%%%%%%%%%%%%%%%%%%%%%%%%%

It is common knowledge that the stratospheric ozone layer is
very important for sustaining life on Earth -- the ozone layer
protects life on Earth from the harmful and damaging ultraviolet
solar radiation. Ozone in the lower atmosphere, or troposphere,
acts as a pollutant but is also an important greenhouse gas.
Ozone is not emitted directly by any natural source. However,
tropospheric ozone is formed under high ultraviolet radiation
flux conditions from natural and anthropogenic emissions of
nitrogen oxides (NO$_x$) and volatile organic compounds (VOCs).
Satellite remote sensing is used to understand and quantify key
processes in the global ozone budgets. Nowadays satellite
observations are readily available for total ozone column and
atmospheric ozone profiles. Nevertheless, ground based
monitoring is important to validate and to complement
space-based measurements and to clarify local/regional specific
sources and sinks of this gas. Such ground based data can assist
to derive the dynamical behavior of air pollution from space and
ground-based observations and to check compliance to the
pollutants transport models. They will also aid to the
development of an environmental policy, in particular policies
on greenhouse gases, on a local and regional scale.

Our first attempts to obtain total ozone columns from FTIR
direct sun observations in the Main Astronomical Observatory
were successful \cite{12} and allows to us to submit the
proposal on OMI validation, which was accepted.

%%%%%%%%%%%%%%%%%%%%%%%%%%%%%%%%%%%%%%%%%%%%%%%%%%%%%%%%%%%%%%%%%%%%%%%%%%%%
\section{OMI satellite observations}
%%%%%%%%%%%%%%%%%%%%%%%%%%%%%%%%%%%%%%%%%%%%%%%%%%%%%%%%%%%%%%%%%%%%%%%%%%%%

The Dutch-Finnish Ozone Monitoring Instrument (OMI) \cite{5,7}
aboard the NASA Earth Observing System (EOS) Aura satellite
\cite{11} is a compact nadir viewing, wide swath,
ultraviolet-visible (270--500~nm) hyperspectral imaging
spectrometer that provides daily global coverage with high
spatial and spectral resolution. The Aura orbit is
sun-synchronous at 705 km altitude with a 98 degrees inclination
and ascending node equator-crossing time roughly at 13:45. OMI
measures backscattered solar radiance in the dayside portion of
each orbit and solar irradiance near the northern hemisphere
terminator once per day. The OMI satellite data products are
derived from the ratio of Earth radiance and solar irradiance.
At the time of writing the OMI TOMS \cite{2,5,8} and OMI DOAS
\cite{2,8,14} total ozone column estimates are publicly
available from the NASA DISC systems. The OMI-TOMS algorithm is
based on the TOMS V8 algorithm that has been used to process
data from a series of four TOMS instruments flown since November
1978. This algorithm uses measurements at 4 discrete 1~nm wide
wavelength bands centered at 313, 318, 331 and 360~nm.

The OMI-DOAS algorithm \cite{14} takes advantage of the
hyper-spectral feature of OMI. It is based on the principle of
Differential Optical Absorption Spectroscopy (DOAS) \cite{9}.
The algorithm uses $\approx25$~OMI measurements in the
wavelength range 331.1~nm to 336.6~nm, as described in
\cite{14}. The key difference between the two algorithms is that
the DOAS algorithm removes the effects of aerosols, clouds,
volcanic sulfur dioxide, and surface effects by spectral fitting
while the TOMS algorithm applies an empirical correction to
remove these effects. In addition, the TOMS algorithm uses a
cloud height climatology that was derived using infrared
satellite data, while the DOAS algorithm uses cloud information
derived from OMI measurements in the 470~nm O$_2$--O$_2$ absorption
band. The two algorithms also respond to instrumental errors
very differently. Validation is key to quantify and understand
these differences as a function of measurement geometry, season
and geolocation.

%%%%%%%%%%%%%%%%%%%%%%%%%%%
\section{Ground based ftir observations}
%%%%%%%%%%%%%%%%%%%%%%%%%%%%%

Ground based FTIR observations are performed with a Fourier
Transform Infra-Red (FTIR) spectrometer, model ``Infralum FT
801'', which was modernized for the task of monitoring the
atmosphere by direct sun observations \cite{3}. The main
advantage of this device is its small size and small sensitivity
of the optical arrangement to vibrations. The working spectral
range of the FTIR spectrometer is 2--12 microns
(800--5000~cm$^{-1}$) with the highest possible spectral
resolution of about 1.00~cm$^{-1}$. Following the modernization
of our spectrometer and upadating the software for the initial
treatment of the registered spectra in 2006 , the system now
allows us to average 2--99 individual spectra during the
observation period. We averaged 4 single spectra as was
recommended by the developers of the spectrometer device to
avoid a degradation of the averaged spectrum due to the
recording of atmospheric instabilities at longer exposure times.
Our averaged spectra have signal-to-noise ratios S/N of
150--200. We registered 3--4 averaged spectra during 2--3
minutes of recording time.

Prior to further treatment of the observed spectra we checked
the repeatability of these 3--4 spectra and choose the spectrum
with the best signal-to-noise ratio S/N to be fitted with the
model spectra.

%%%%%%%%%%%%%%%%%%%%%%%%%%%%%
\section{Modtran spectra modeling and analysis}
%%%%%%%%%%%%%%%%%%%%%%%%%%%%%
%%%%%%%%%%%%%%%%%%%%%%%%%%%%%
The column amounts of  ozone (O$_3$) molecules are recovered by
using the radiation transfer codes MODTRAN3 and MODTRAN4, a
moderate resolution model of transmission \cite{1}. These codes
are widely applied to the interpretation of ground based,
airborne and spaceborne (satellite) observations of spectra of
the Earth's atmosphere. The codes calculate atmospheric
transmission and reflection of electromagnetic radiation with
frequencies from 0 up to 50000~cm$^{-1}$. The model uses a
spherical source function for the light originating from the Sun
and scattered from the Moon, and standard model atmospheres and
user specified atmospheric profiles of gases, aerosols, clouds,
fogs and even rain. It uses a two-parameter (temperature and
pressure) model of molecular absorption bands, which is
calculated on the basis of a large array of previously
accumulated data of spectral lines stored in the HITRAN
database. Here we use absorption cross-section data for 12 light
molecules (H$_2$O, CO$_2$, O$_3$, CO, CH$_4$, O$_2$, NO, SO$_2$,
NO$_2$, N$_2$O, NH$_4$ and HNO$_3$), for heavy molecules -- CFC
(9 molecules) and for ClONO$_2$, HNO$_4$, CCl$_{4}$ and
N$_2$O$_5$. The calculations are carried out only in an local
thermal equilibrium (LTE) approximation for the moderate
spectral resolution (2~cm$^{-1}$) which just corresponds to our
observed Fourier spectra. The Band Model parameters were
re-calculated by us on the base of HITRAN-2004 according to the
paper of \cite{1}. Measurements of surface ozone concentrations
by the collocated ozonometer together with satellite remote
sensing data from the Atmospheric Infrared Sounder Instrument
(AIRS - http://disc.gsfc.nasa.gov/AIRS/) aboard the NASA
EOS-Aqua platform and the Microwave Limb Sounder ((MLS,
http://avdc.gsfc.nasa.gov/Data/Aura/) aboard the NASA EOS-Aura
platform were used for the construction of atmospheric ozone,
temperature and water vapor input profiles for the MODTRAN4.3
code. For the analysis of the 2006 FTIR observations we used MLS
version 1.5 data, which then had a preliminary character. We
modified the shape of the MLS stratospheric ozone profile to
obtain a better fit to the MODTRAN4.3 model output and to our
FTIR spectra of the ozone band around 9.6 microns \cite{13}.

Fortunately, in 2007 the all new and more precise version v2.2
of MLS data became available, that allows us to develop a new
approach to the analysis: we now modified the input tropospheric
ozone profile and we only scaled the stratospheric ozone
profiles of Aura-MLS v2.2 data within 2--5\% (declared precision
of these data) without any modification to its shape. The
tropospheric part of the input (a priori) ozone profile was
constructed from surface ozone measurement and the TEMIS
climatological (monthly averaged) ozone atmospheric profiles
\cite{4}, which were downloaded from the TEMIS-KNMI website. In
this way we tried to obtain the best possible fit of the model
computed spectra to the FTIR observed spectral band on 9.6
microns. Aura-TES data available from the AVDC website were also
used if they were available for observational days.

To modify the tropospheric ozone profiles we used a smooth function
determined between the $J1$ and $J2$ points of altitude in the model
atmosphere. For any $J$ point of the model we then adopt:
 $$
 x(B)=(J-J1)/(J2-J1),
 $$ \noindent
 then
 $$
 P_J=P0_J(1+B(\sin x))^a
 $$ \noindent
determines the shape of the correction function, $a$
and $B$ determine the amplitude of changes of input tropospheric
ozone profile, where $B> -1$ and $a > 0$. Using the MODTRAN4
code we compute a grid of the theoretical spectra. To determine
the best fit parameters, we compare the observed and computed
spectra following a two-step optimization procedure.

Firstly, we determined the best fit to observed water vapor
lines in the spectral region 800--1240~cm$^{-1}$, i.e., here we
exclude the ozone band from the analysis. Secondly, we fit the
observed spectrum around the 9.6 micron ozone band with the grid
of calculated ozone bands including the previously determined
best atmospheric water profile. Hence we determine tropospheric
ozone profiles, total and tropospheric ozone column from the
best fit of the modeled and observed ozone band spectra, where
we
%---------------------------------------------------------------
%        1                                        One column figure
%-----------------------------------------------------------
   \begin{figure}
   \centering
  \includegraphics[width=15cm]{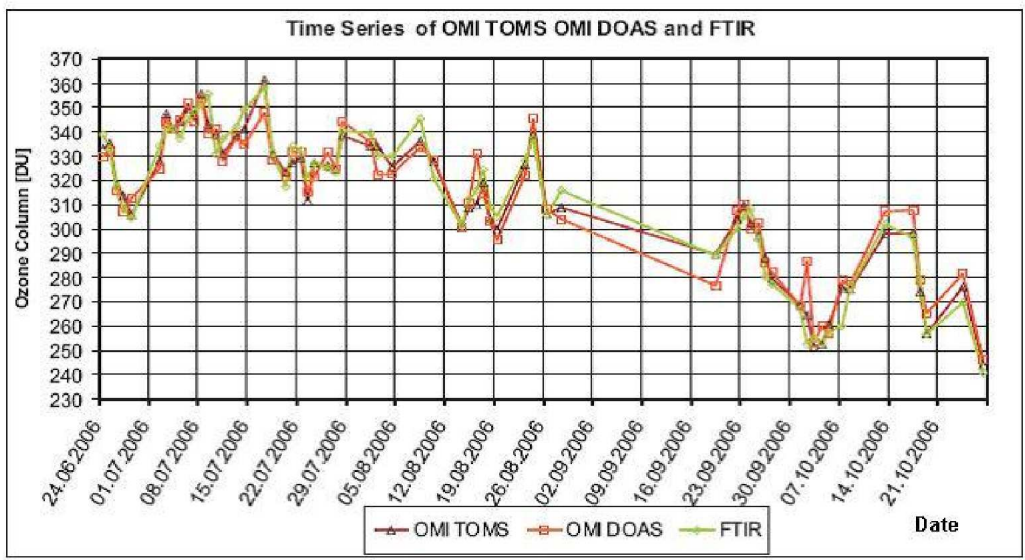}
  \includegraphics[width=13cm]{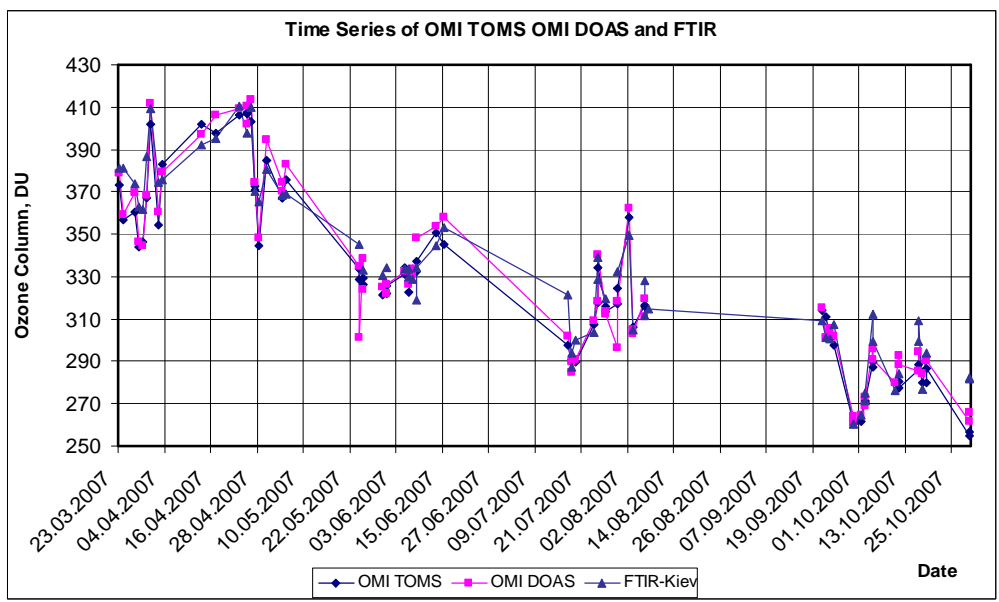}
      \caption{{\small  a) Time series of the OMI total ozone column
and the ground based FTIR total ozone data of 2006 for the
ground site of Kiev (MAO). Average difference of satellite minus
ground based amounts to 0.37 DU and -0.25 DU for OMI-DOAS and
OMI-TOMS respectively, with a 8.77 DU and 5.37 DU standard
deviation (1.11 DU and 0.68 DU standard errors). b)Time series
of the OMI total ozone and the ground based FTIR total ozone
data of 2007 for Kiev (MAO).  Average difference of satellite
minus ground based amounts to -0.24 DU and -4.17 DU for OMI-DOAS
and OMI-TOMS respectively, with 10.50 DU and 10.73 DU standard
deviations (1.31 DU and 1.35 DU standard errors)}.
              }
         \label{Fig1}
   \end{figure}\noindent
included the unaltered Aura-MLS stratospheric profiles.

Figure 1 (a,b) presents the comparison of our FTIR total ozone
column values with the OMI-DOAS and OMI-TOMS satellite total
ozone column data for the year 2006 and 2007. On average the
difference of satellite and ground based observations in 2006
amounts to 0.37~DU and -0.25~DU for OMI-DOAS and OMI-TOMS
respectively, with a 8.77~DU and 5.37~DU standard deviation
(1.11~DU and 0.68~DU standard errors). For 2007  average
difference of satellite minus ground based amounts to -0.24 DU
and -4.17 DU for OMI-DOAS and OMI-TOMS respectively, with 10.50
DU and 10.73 DU standard deviations (1.31 DU and 1.35 DU
standard errors).

%   \begin{figure}
%   \centering
%  \includegraphics[width=13cm]{fig11.eps}
%      \caption{Time series of the OMI total ozone and the ground based FTIR
%total ozone data of 2007 for Kiev (MAO).  Average difference of satellite minus
%ground based amounts to -0.24 DU and -4.17 DU for OMI-DOAS and OMI-TOMS
%respectively, with 10.50 DU and 10.73 DU standard deviations
%(1.31 DU and 1.35 DU standard errors).
%              }
%         \label{Fig1b}
%   \end{figure}\noindent
%

%______________________________________________________________

%         2                                       One column figure
%-----------------------------------------------------------
   \begin{figure}
   \centering
  \includegraphics[width=8.cm]{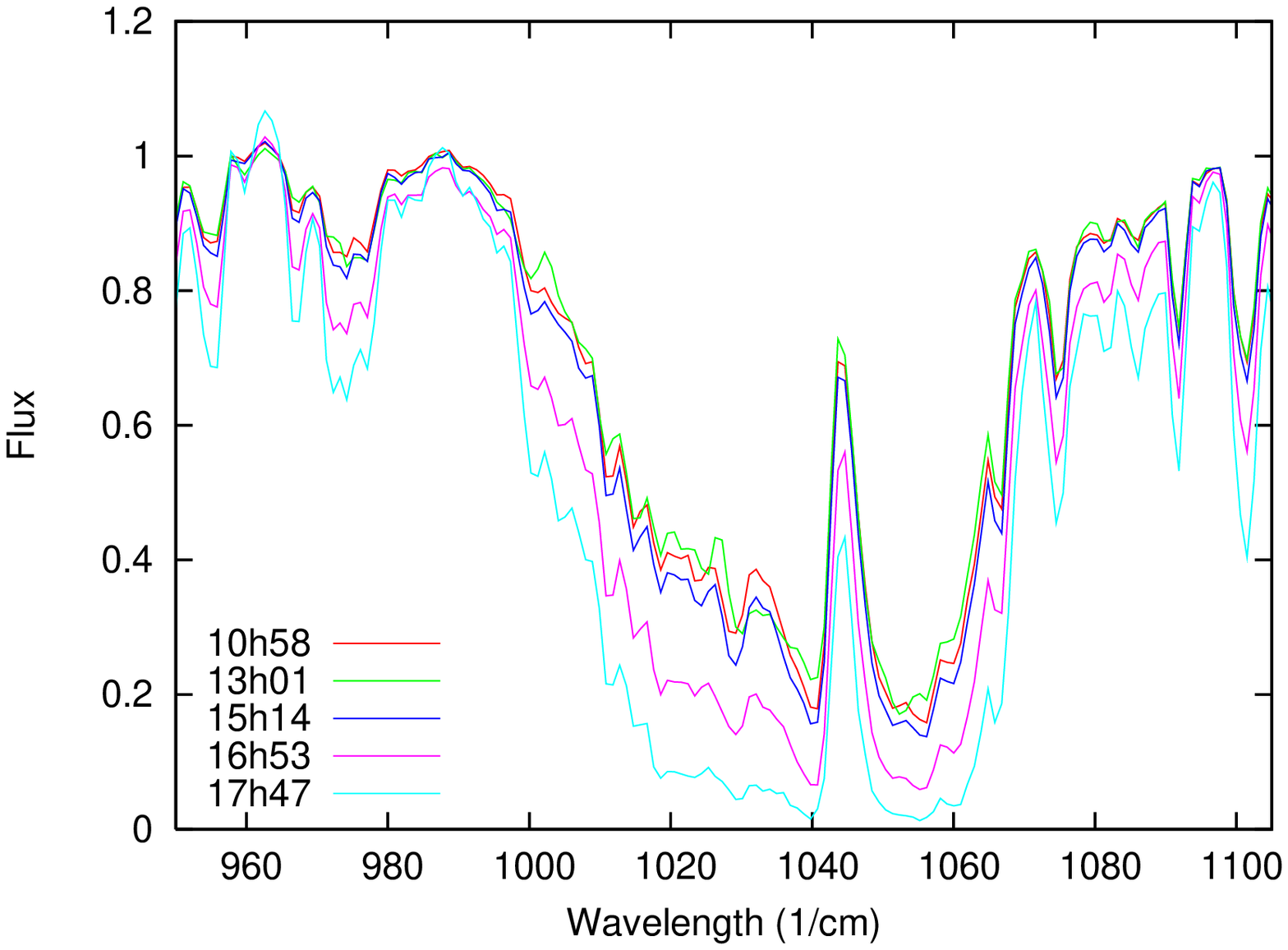}
  \includegraphics[width=8.cm]{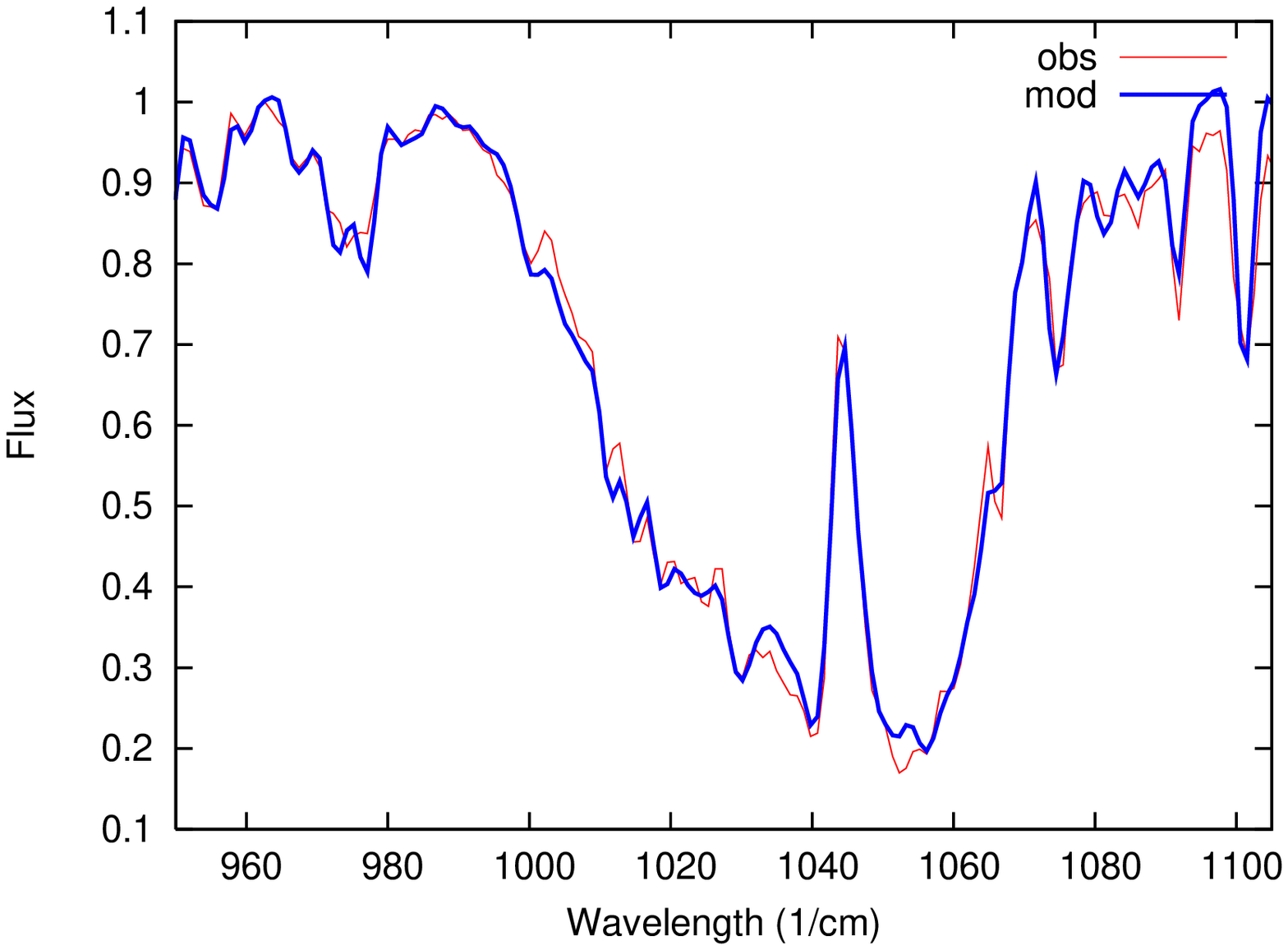}
      \caption{{\small  The observed FTIR spectra of the 9.6 micron
ozone band for the 29th of September 2007 (29.09.07) (left) and
the comparison of the observed FTIR spectra and modeled MODTRAN
4 spectra following the procedure for best fitting for the
observation at 13~h~01~min local time on this day.}
              }
         \label{Fig2}
   \end{figure}
%
%______________________________________________________________
%          3                                      One column figure
%----------------------------------------------------------- S_vib
   \begin{figure}
   \centering
 \includegraphics[width=8cm]{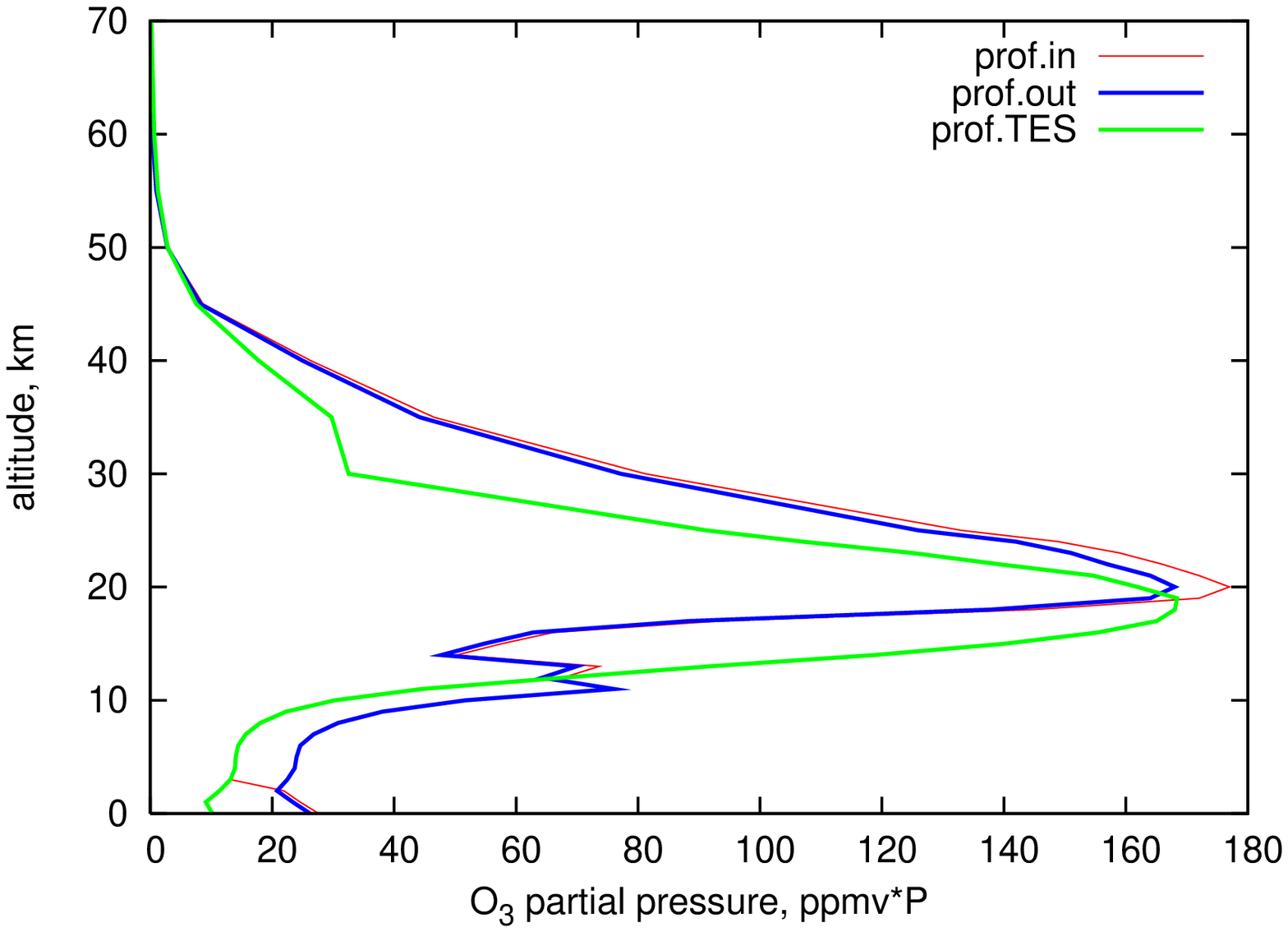}
 \includegraphics[width=8cm]{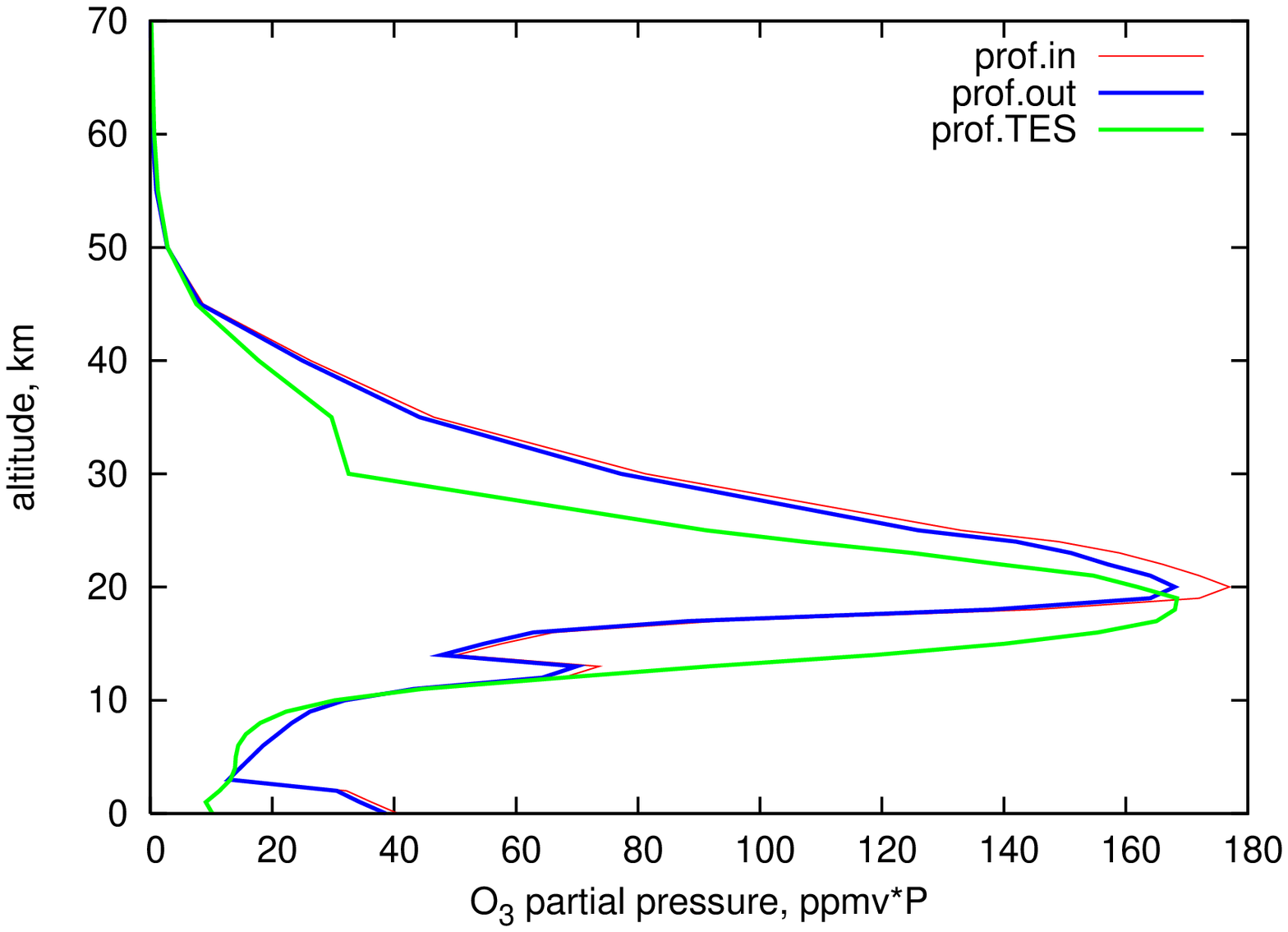}
 \includegraphics[width=8cm]{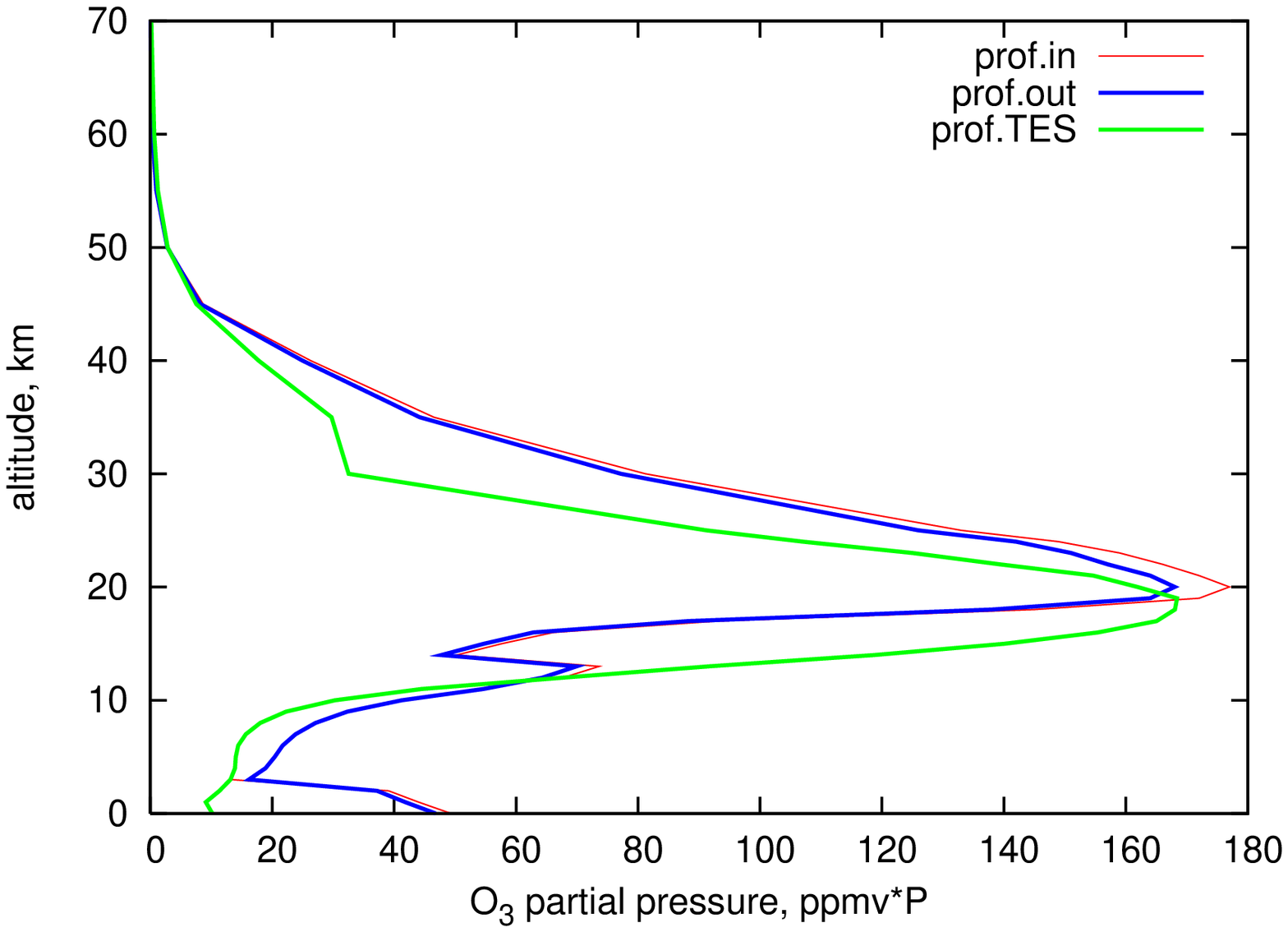}
 \includegraphics[width=8cm]{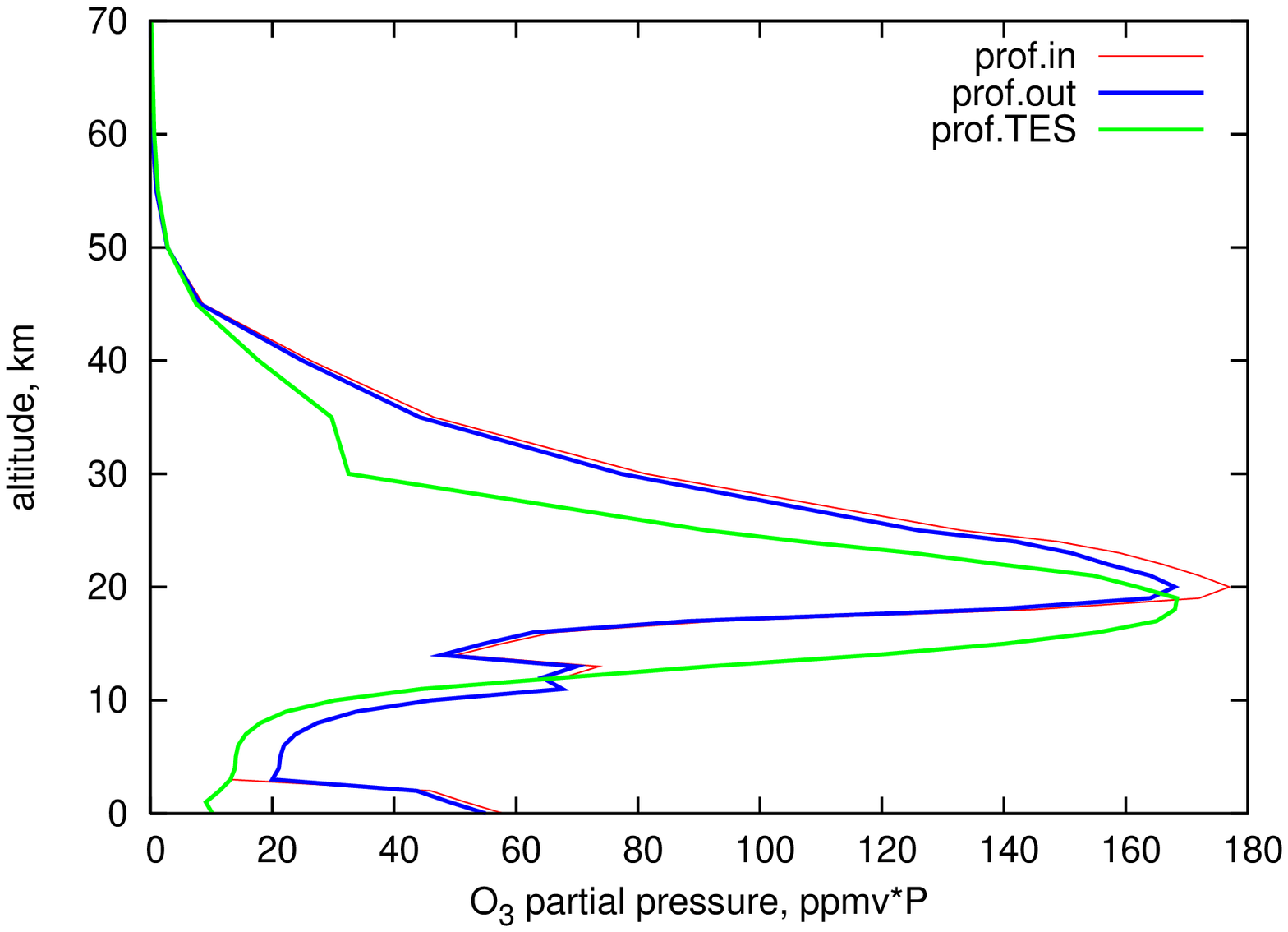}
      \caption {{\small  The retrieved ozone atmospheric profiles
for the 28th of March 2007 (28.03.07), recorded at 08~h~54~min
and 10~h~47~min (upper figures) local time, and 13~h~12~min and
18~h~21~min (lower figures) local time. From these figures one
observes the low ozone concentrations in the boundary layer for
the morning observation at 08~h~54~min LT. Here most probably
ozone titration by the nitrogen oxides (NOx) as emitted from
cars during the morning traffic is taking place. From the
10~h~47~min LT observation we see the abatement of tropospheric
ozone, most clearly over the vertical range 2--11~km. The
enhancements of ozone due to the photochemical processes in the
atmosphere are seen in the lower two figures. Our simultaneously
performed surface ozone measurements reflect this dynamics also
with the supportive values 27.3~ppb, 40.2~ppb, 48.8~ppb, and
57.3~ppb recorded for exactly these moments in time. For the
comparison, we also show the Aura-TES ozone vertical profile for
the 28th of March 2007 (28.03.07), which can be considered as
the valid satellite profile in the troposphere only.}
              }
         \label{Fig3}
   \end{figure}

In Figure 2 we show in the left figure all the observed spectra
for the day 29th of September 2007 (29.09.07). We also
demonstrate the best fit of the modeled spectra to the FTIR
spectrum observed at 13~h~01~min local time in  Figure 2
(right).

%______________________________________________________________
%           4                                     One column figure
%-----------------------------------------------------------
   \begin{figure}
   \centering
 \includegraphics[width=8cm]{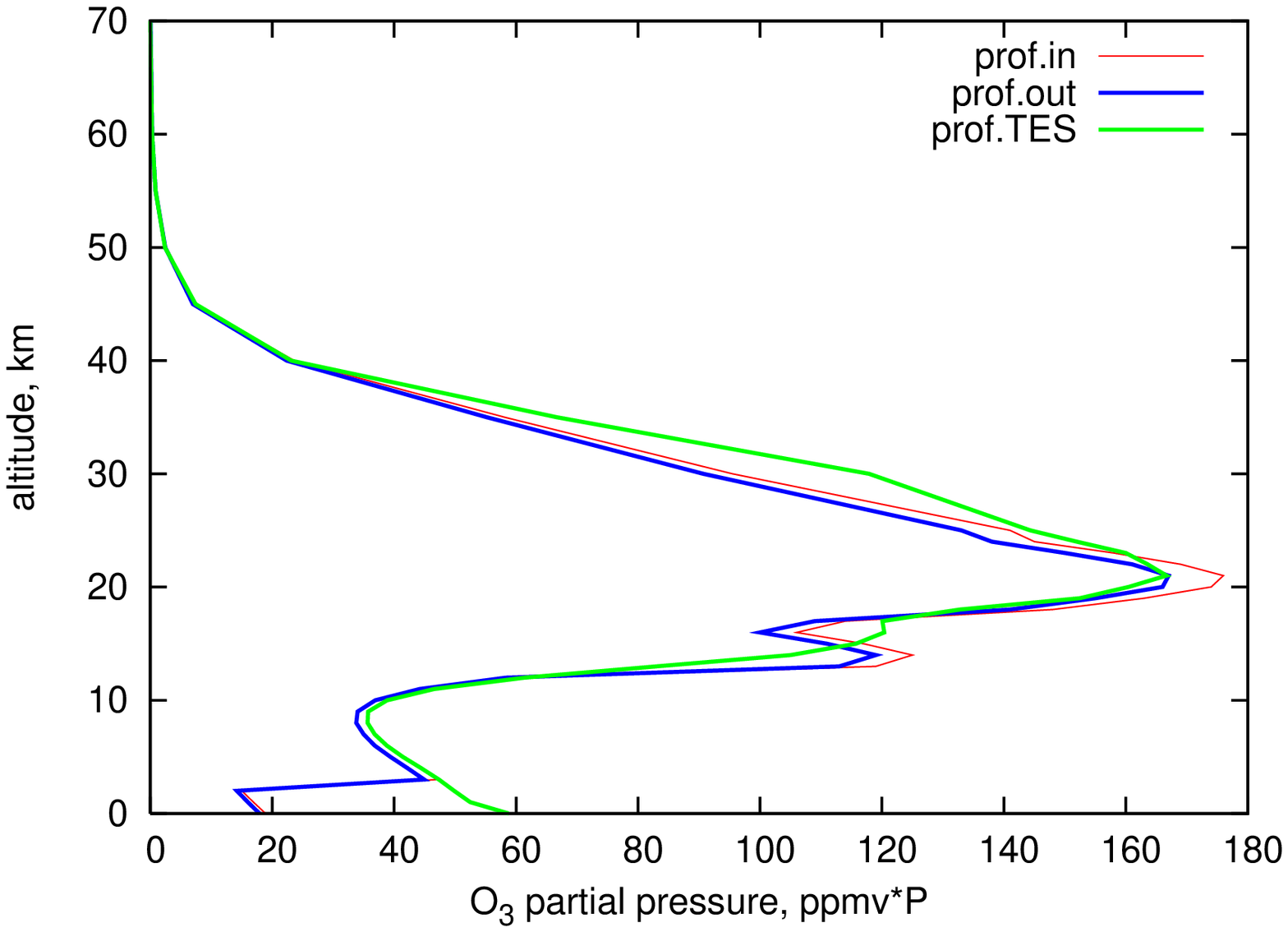}
 \includegraphics[width=8cm]{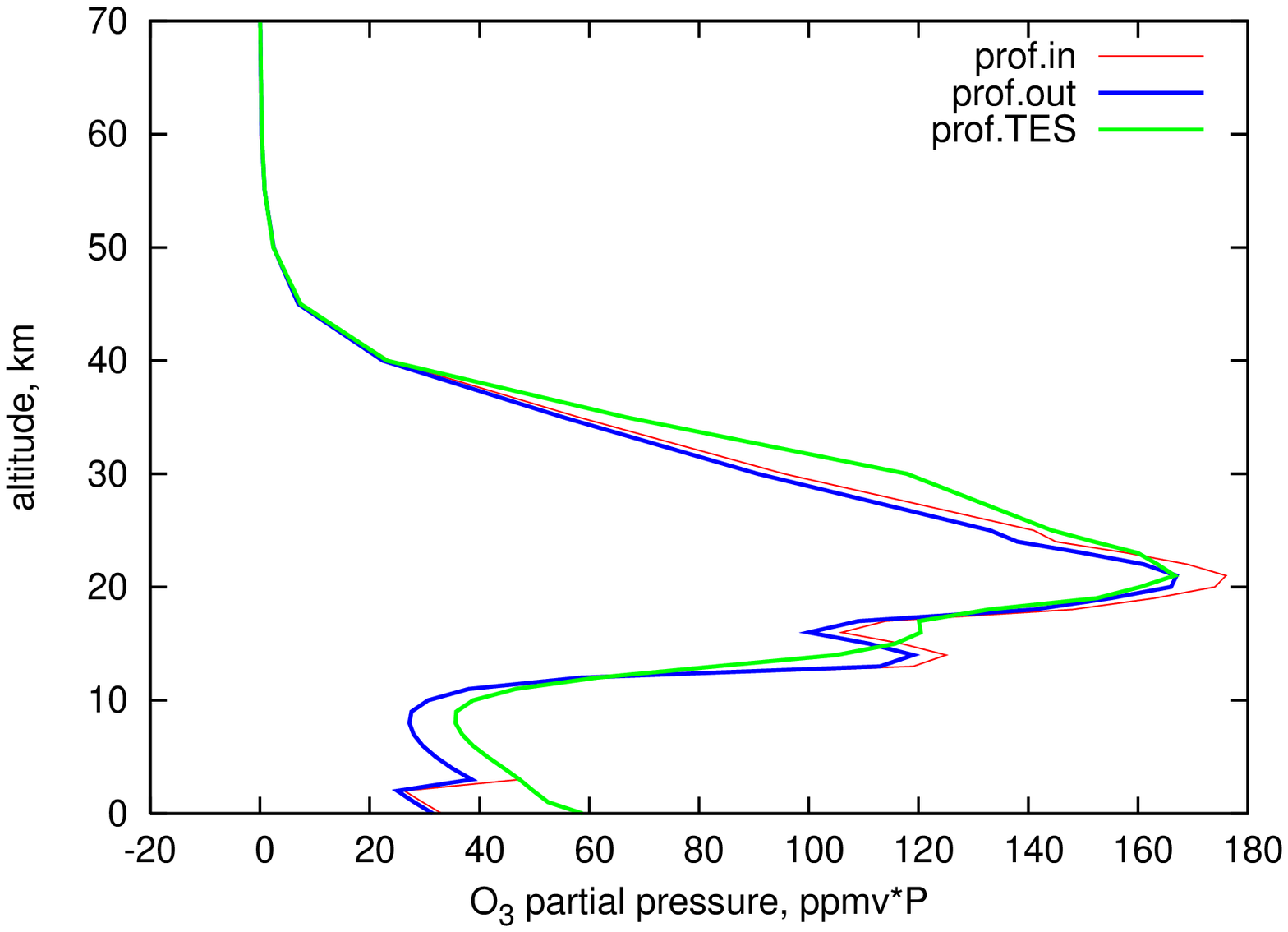}
 \includegraphics[width=8cm]{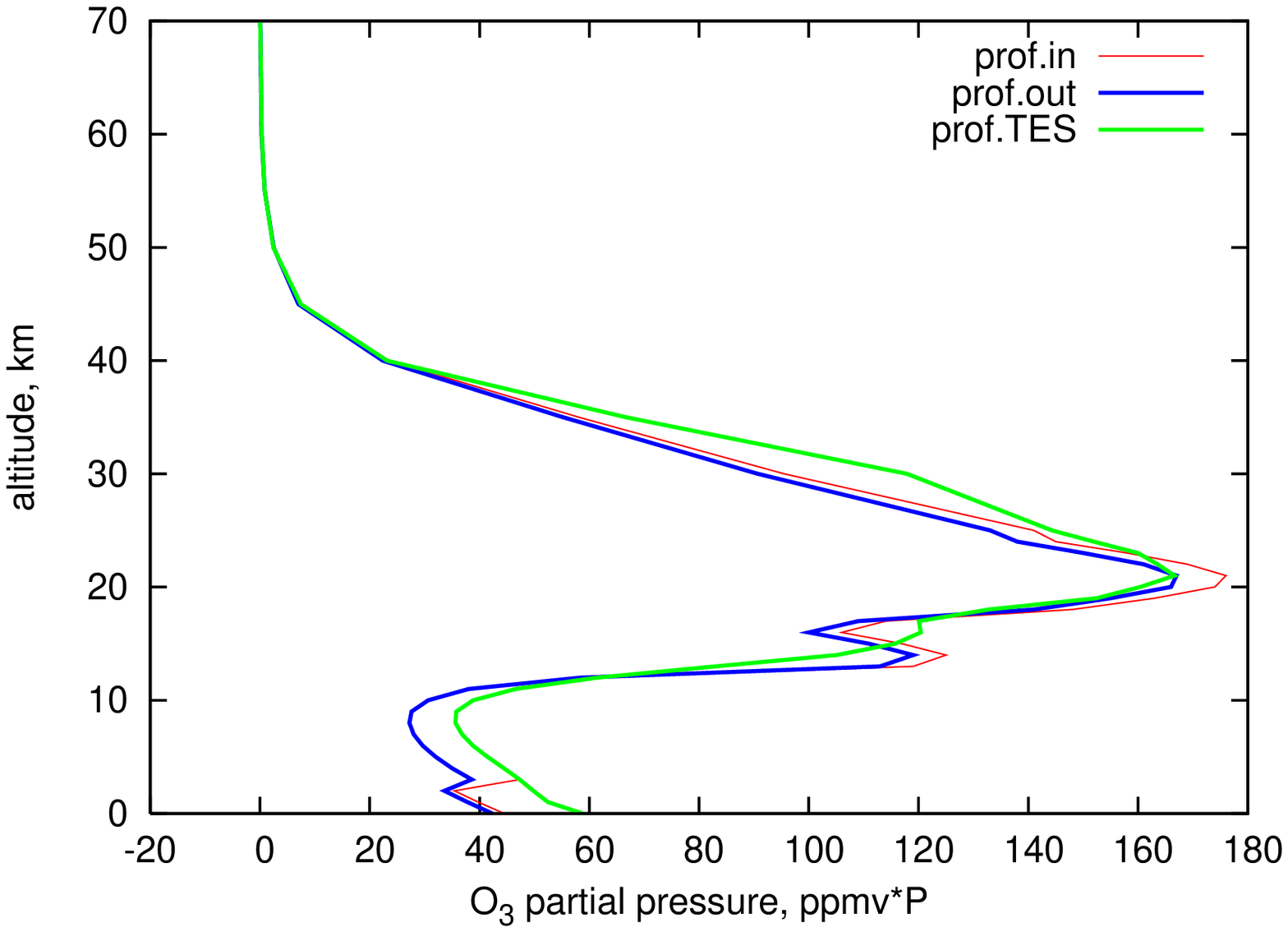}
 \includegraphics[width=8cm]{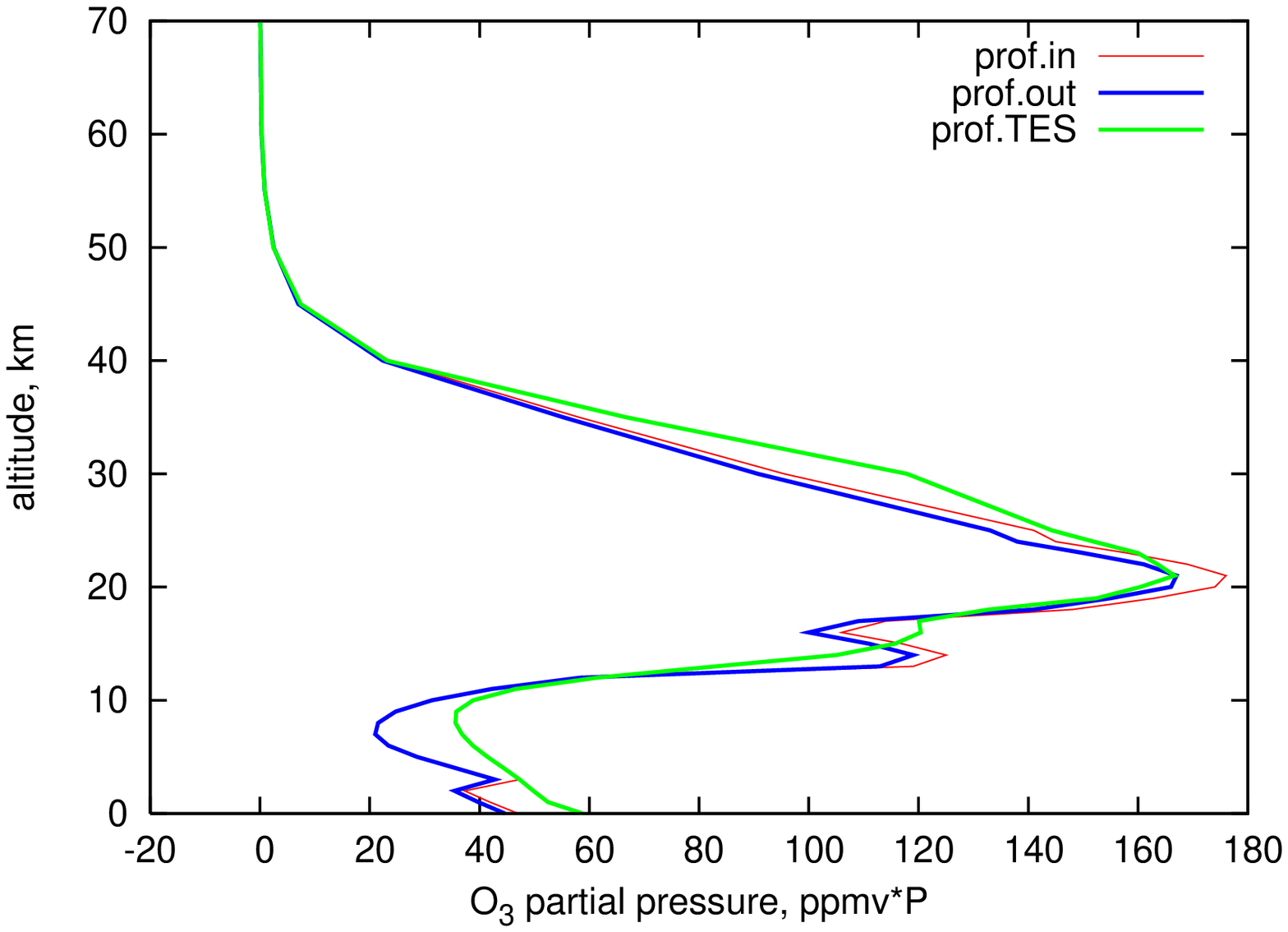}
      \caption{{\small The retrieved ozone atmospheric profiles for
the 23rd of April 2007 (23.04.07) recorded at 09~h~22~min and
11~h~15~min (upper figures) local time, and recorded at
14~h~35~min and 15~h~40~min (lower figures) local time. On this
day the values of both total ozone columns (411.0~DU by FTIR)
and tropospheric ozone columns are very high. Possibly we are
here observing a stratospheric intrusion event as the highest
OMI value of total ozone column in 2007 was 448 DU for the 22nd
of April 2007 (22.04.07),
 the day before.}
              }
         \label{Fig4}
   \end{figure}
%
%______________________________________________________________
%            5                                    One column figure
%-----------------------------------------------------------
   \begin{figure}
   \centering
 \includegraphics[width=8cm]{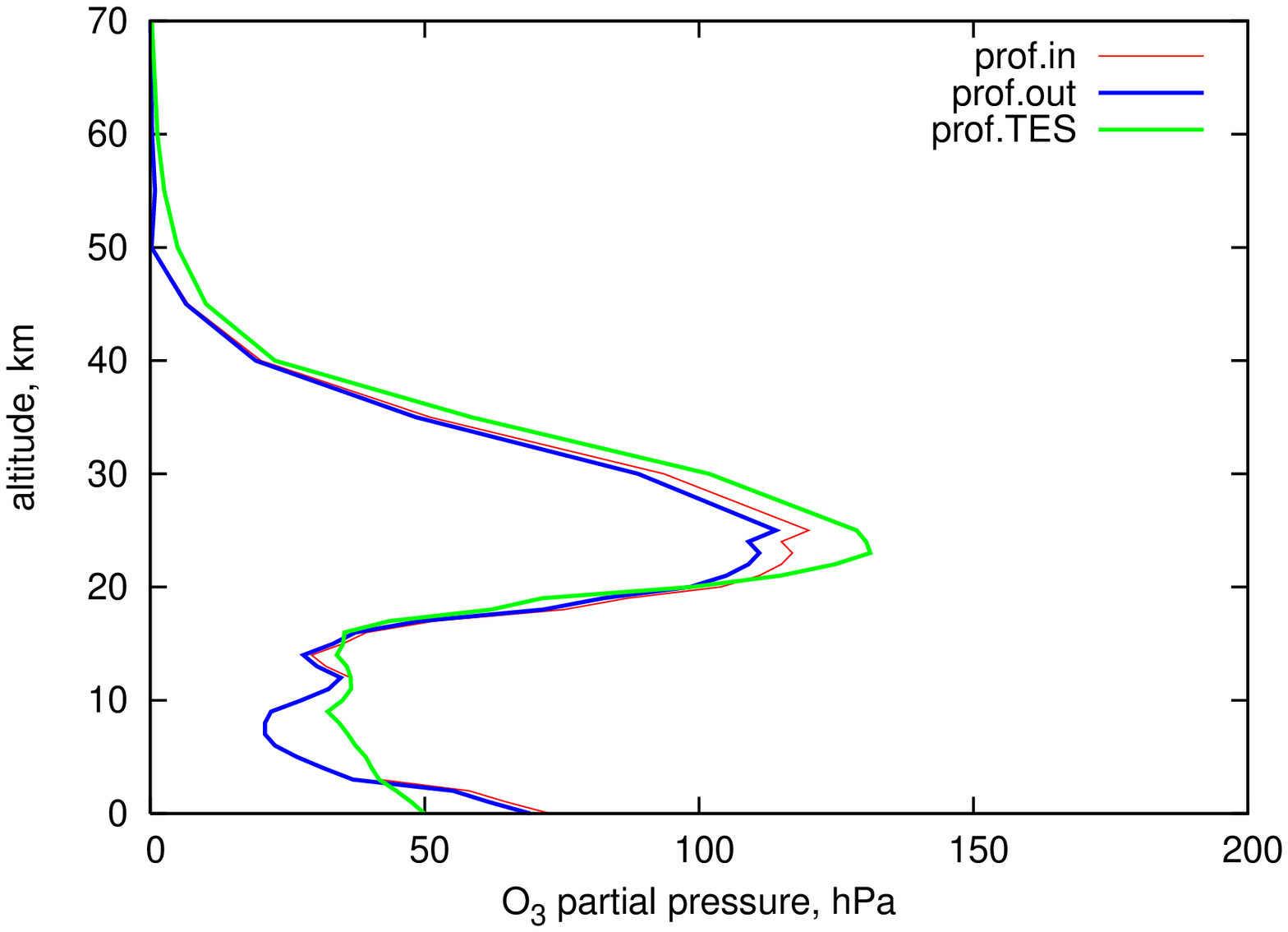}
 \includegraphics[width=8cm]{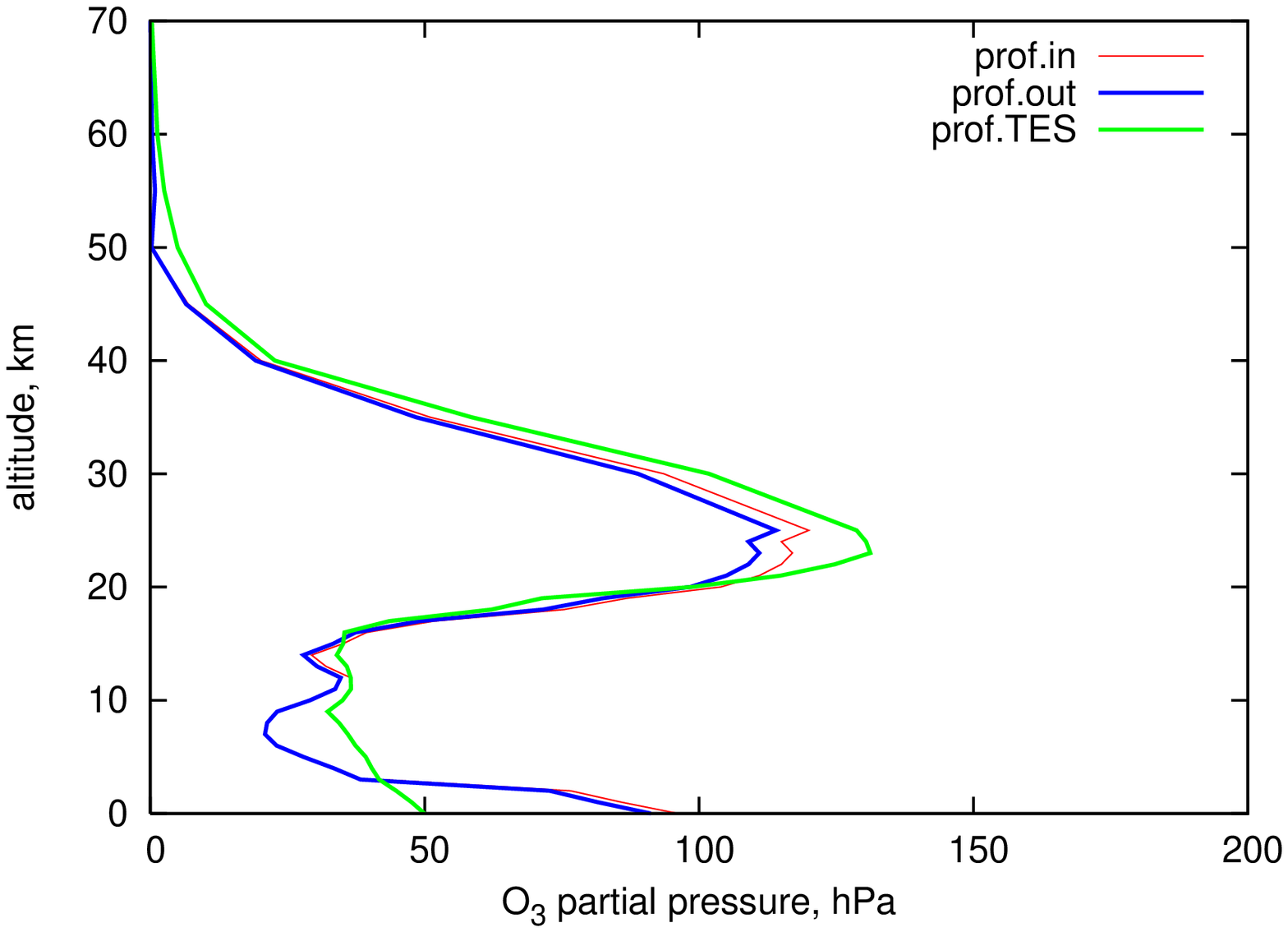}
 \includegraphics[width=8cm]{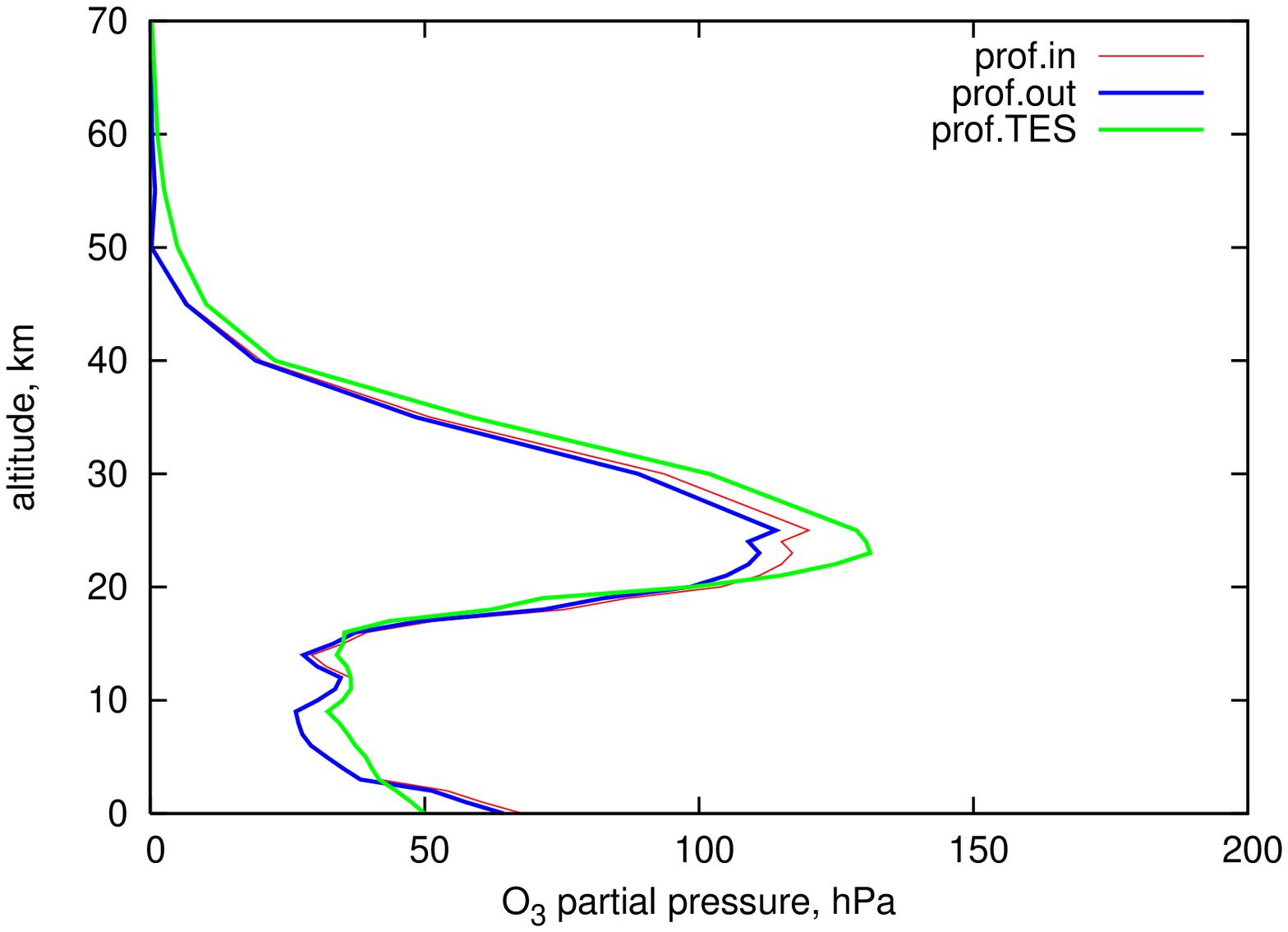}
 \includegraphics[width=8cm]{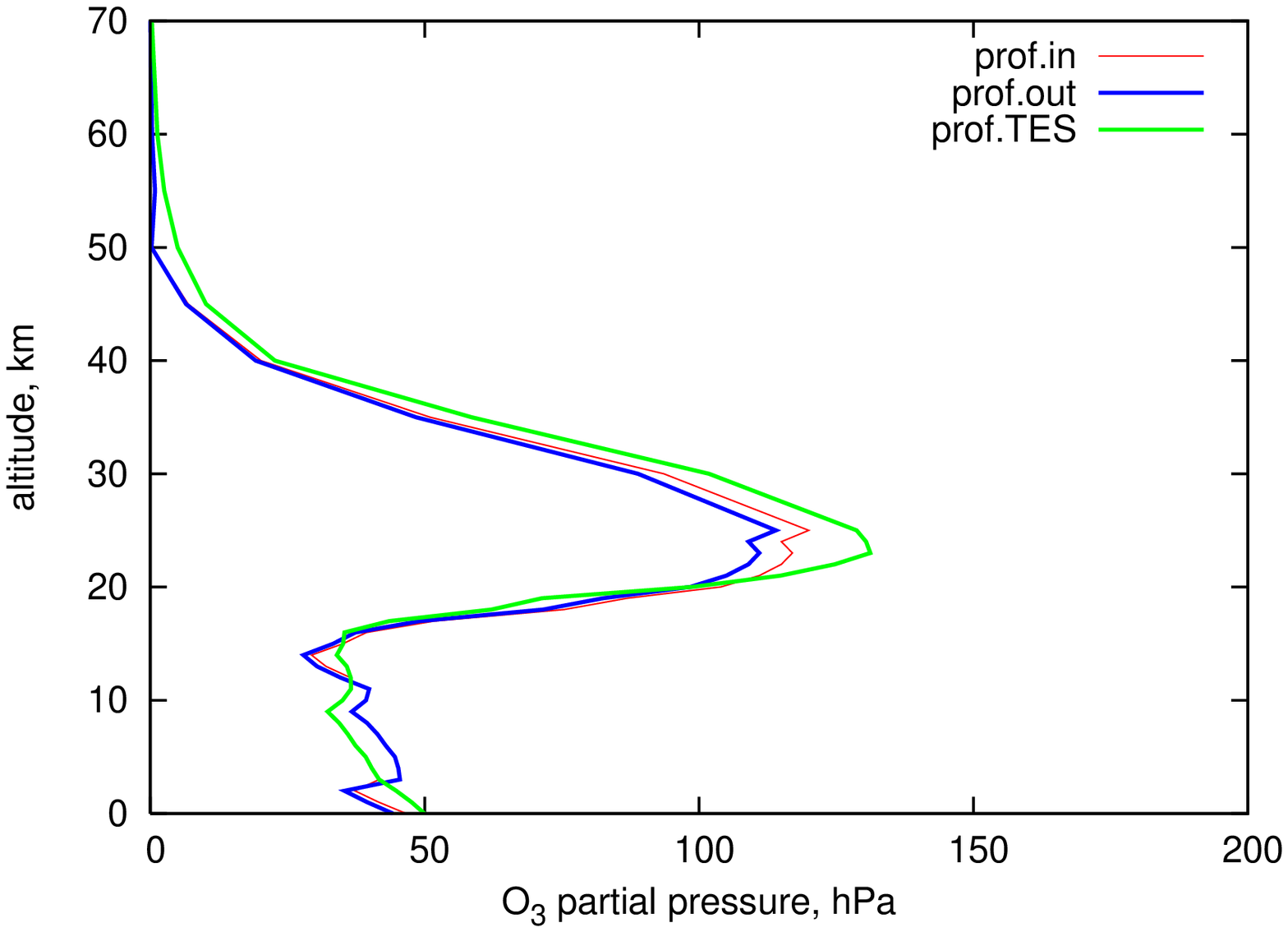}
      \caption{{\small The retrieved ozone atmospheric profiles for
the 18th of July 2007 (18.07.07) recorded at 13~h~35~min and
16~h~10~min (upper figures) local time, and recorded at
17~h~20~min and 19~h~27~min (lower figures) local time. The very
high tropospheric ozone columns and surface ozone concentrations
(see Table 1 for the exact numbers) and their daily dynamics are
characteristic for episodes of strongly enhanced surface and
troposheric ozone due to tropospheric photochemistry. Please
note that on this day the total ozone column is actually rather
low (291.5~DU)}.
              }
         \label{Fig5}
   \end{figure}

The Figures 3-6 present our retrieved ozone profiles for four
representative cases: two spring enhanced ozone episodes, summer
ozone photochemistry episode and autumn low stratospheric ozone
episode. The latter case was observed during 3 days: 29th of
September 2007 (29.09.07), 1st of October 2007 (01.10.07) and
the 2nd of October 2007 (2.10.07). The figures demonstrate the
specifics of each episode and the daily dynamics of tropospheric
ozone variability due to the underlying photochemical processes.

Figure 3 shows the retrieved ozone atmospheric profiles for the
28th of March 2007 (28.03.07), recorded at 08~h 54~min and
10~h~47~min (upper figures) local time, and 13~h~12~min and
18~h~21~min (lower figures) local time. From these figures one
observes the low ozone concentrations in the boundary layer for
the morning observation at 08~h~54~min LT. Here most probably
ozone titration by the nitrogen oxides (NO$_x$) as emitted from
cars during the morning traffic is taking place. From the
10~h~47~min LT observation we see the abatement of tropospheric
ozone, most clearly over the vertical range 2--11~km. The
enhancement of ozone due to the photochemical processes in the
atmosphere are seen in the lower two figures. Our simultaneously
performed surface ozone measurements reflect this dynamics also
with the supportive values 27.3~ppb, 40.2~ppb, 48.8~ppb, and
57.3~ppb recorded for exactly these moments in time. For the
comparison, we also show the Aura-TES ozone vertical profile for
the 28-th of March 2007 (28.03.07), which can be considered as
the valid satellite profile in the troposphere only.

Figure 4 shows the retrieved ozone atmospheric profiles for the
23rd of April 2007 (23.04.07) recorded at 09~h~22~min and
11~h~15~min (upper figures) local time, and recorded at
14~h~35~min and 15~h~40~min (lower figures) local time. On this
day the values of both total ozone columns (411.0~DU by FTIR)
and tropospheric ozone columns are very high. Possibly we are
here observing a stratospheric intrusion event as the highest
OMI value of total ozone column in 2007 was 448~DU for the 22nd
of April 2007 (22.04.07).

Figure 5 shows the retrieved ozone atmospheric profiles for the
18-th of July 2007 (18.07.07) recorded at 13~h~35~min and 16~h
10~min (upper figures) local time, and recorded at 17~h 20~min
and 19~h~27~min (lower figures) local time. The very high
tropospheric ozone columns and surface ozone concentrations
(see, for the exact numbers) and their daily dynamics are
characteristic for episodes of strongly enhanced surface and
troposheric ozone due to tropospheric photochemistry.

%_____________________________________________________________
%                                          Table with footnotes
%-------------------------------------------------------------
%
\begin{table}
\begin{minipage}[t]{\columnwidth}
\caption{Total ozone columns and tropospheric ozone columns for
some days of 2007.} \label{tab1} \centering \medskip
\renewcommand{\footnoterule}{}  % to avoid a line before footnotes
\small
\begin{tabular}{|c|c|c|c|c|c|c|c|c|}
\hline \hline Date& Time,     &  ZSA,    & TOC, &
OMI-TOMS,&OMI-DOAS,&Tr.OC, DU,&Surface&Htrop,
\\ ~ &h, min & grad &DU& DU&DU&our, TES& O$_3$, ppb&km  \\ \hline
28.03.07 & 8 54   & 70.434 &364.24  & ~    & ~     & 47.15&
27.3& 12.0\\ ~        & 10 47  & 58.459 &363.57  & ~    & ~
& 36.06& 40.2& ~\\ ~        & 13 12  & 47.469 &361.39  & 344.2&
356.0 & 44.13& 48.8& ~\\ ~        & 14 46  & 52.131 &363.94  &
353.2& 363.2 & 46.72& 65.9& ~\\ ~        & 16 51  & 67.169
&363.54  & ~    & ~     & 46.33& 64.0& ~\\ ~        & 17 51  &
76.192 &359.91  & ~    & ~     & 43.54& 56.5& ~\\ ~        & 18
21  & 80.375 &366.27  & ~    & ~     & 44.93& 57.3& ~\\ \hline
23.04.07 &9 22    & 57.622 &411.01  & ~    & ~     & 48.06&
18.7& 12.5\\ ~        &11 15   & 43.200 &410.30  & ~    & ~
& 47.34& 32.8& ~\\ ~        &14 35   & 42.879 &410.27
&412.0\footnote{22.04.07 OMI total column value = 448 DU} &
414.5 & 47.30& 44.1& ~\\ ~        &15 40   & 50.375 &409.54
&414.7 & 417.6 & 46.57& 46.5& ~\\ \hline
 9.06.07 & 6 39   & 75.28 &348.37  & ~    & ~     & 38.40&20  & 12.0\\
~        & 8 44   & 55.66 &341.53  & ~    & ~     & 31.70&22  &
~\\ ~        & 11 56  & 29.93 &346.05  & 347.6& 349.0 &
35.47&42.8& ~\\ ~        & 16 08  & 45.92 &352.76  & ~    & ~
& 36.38&51  & ~\\ ~        & 17 53  & 62.42 &349.56  & ~    & ~
& 39.56&57  & ~\\ \hline 14.06.07 & 6 52  & 73.20 &355.54  & ~
& ~     & 42.9 &14& 12.0\\ ~        & 7 05  & 71.21 &351.04  & ~
& ~     & 44.75&13& ~\\ ~        & 9 05  & 52.25 &352.81  & ~
& ~     & 39.81&15& ~\\ ~        & 12 06 & 28.96 &348.72  &
347.6& 349.6 & 42.86&46& ~\\ ~        & 17 45 & 60.75 &357.36  &
~    & ~     & 44.76&50& ~\\ \hline 18.07.07 & 13 35 & 29.93
&287.12  & 291.5& 289.6 &44.32&72& 12.6\\ ~        & 14 52 &
36.16 &294.07  & ~    & ~     &51.27&85& ~\\ ~        & 16 10 &
46.62 &290.91  & ~    & ~     &49.37&95& ~\\ ~        & 17 20 &
58.23 &294.39  & ~    & ~     &51.60&67& ~\\ ~        & 18 15 &
66.19 &292.85  & ~    & ~     &50.09&58& ~\\ ~        & 19 27 &
77.39 &296.60  & ~    & ~     &53.80&46& ~\\ ~        & ~     &
~     &~       & ~    & ~     &53.55\footnote{TESL3 tropospheric
ozone column}&~ & ~\\ \hline 29.09.07 & 10 35  & 57.722 &269.21
& ~    & ~     & 29.96&13.0& 13.0\\ ~        & 13 01  & 52.756
&260.34  & 261.2& 263.9 & 32.64&29.0& ~\\ ~        & 15 14  &
61.211 &260.38  & ~    & ~     & 32.31&39.0& ~\\ ~        & 16
37  & 71.676 &261.44  & 260.2& 260.9 & 33.73&40.0& ~\\ ~
& 17 47  & 82.059 &266.62  & ~    & ~     & 38.92&35.0& ~\\
\hline
 1.10.07 & 8 08   & 79.704 &271.75  & ~    & ~     & 30.31&8.0& 12.5\\
~        & 9 49   & 65.672 &261.95  & ~    & ~     & 28.09&18.0&
~\\ ~        & 13 22  & 53.971 &264.68  & 261.7& 264.9 &
30.69&40.0& ~\\ ~        & 16 21  & 70.201 &271.41  & ~    & ~
& 37.42&45.0& ~\\ ~        & 17 41  & 81.844 &277.23  & ~    & ~
& 43.24&39.0& ~\\ \hline
 2.10.07 & 8 31   & 76.545 &279.16  & ~    & ~     & 40.43&8.0& 12.5\\
~        & 9 43   & 66.709 &276.51  & ~    & ~     & 37.78&12.0&
~\\ ~        & 12 58  & 53.897 &271.42  & 270.9& 269.1 &
34.93&43.0& ~\\ ~        & 15 20  & 63.019 &274.80  & ~    & ~
& 36.08&47.0& ~\\ ~        & ~      & ~      &~       & ~    & ~
& 39.19$^{\it b}$&~   & ~\\ \hline

\end{tabular}
\end{minipage}
\end{table}

Please note that on this day the total ozone column is actually
rather low (291.5~DU).Finally, in Figure 6 we show the retrieved
atmospheric ozone profiles for the 1st of October 2007
(01.10.07) recorded at 8~h~8~min and 9~h~49~min (upper figures)
local time and recorded at 16~h 21~min and 17~h~41~min (lower
figures) local time. Please note that on this day the FTIR total
ozone column is rather low: only 262~DU. Nevertheless, we can
see the daily dynamics of tropospheric ozone: in the morning
ozone titration by NO$_x$ is taking place leading to rather high
ozone concentrations later in the afternoon. Unfortunately, for
this day Aura-TES data are absent and hence the tropospheric
part of the input ozone profile for the MODTRAN modeling process
was constructed on the basis of the TEMIS monthly averaged
climatological data.

%______________________________________________________________
%             6                                   One column figure
%-----------------------------------------------------------
   \begin{figure}
   \centering
 \includegraphics[width=8cm]{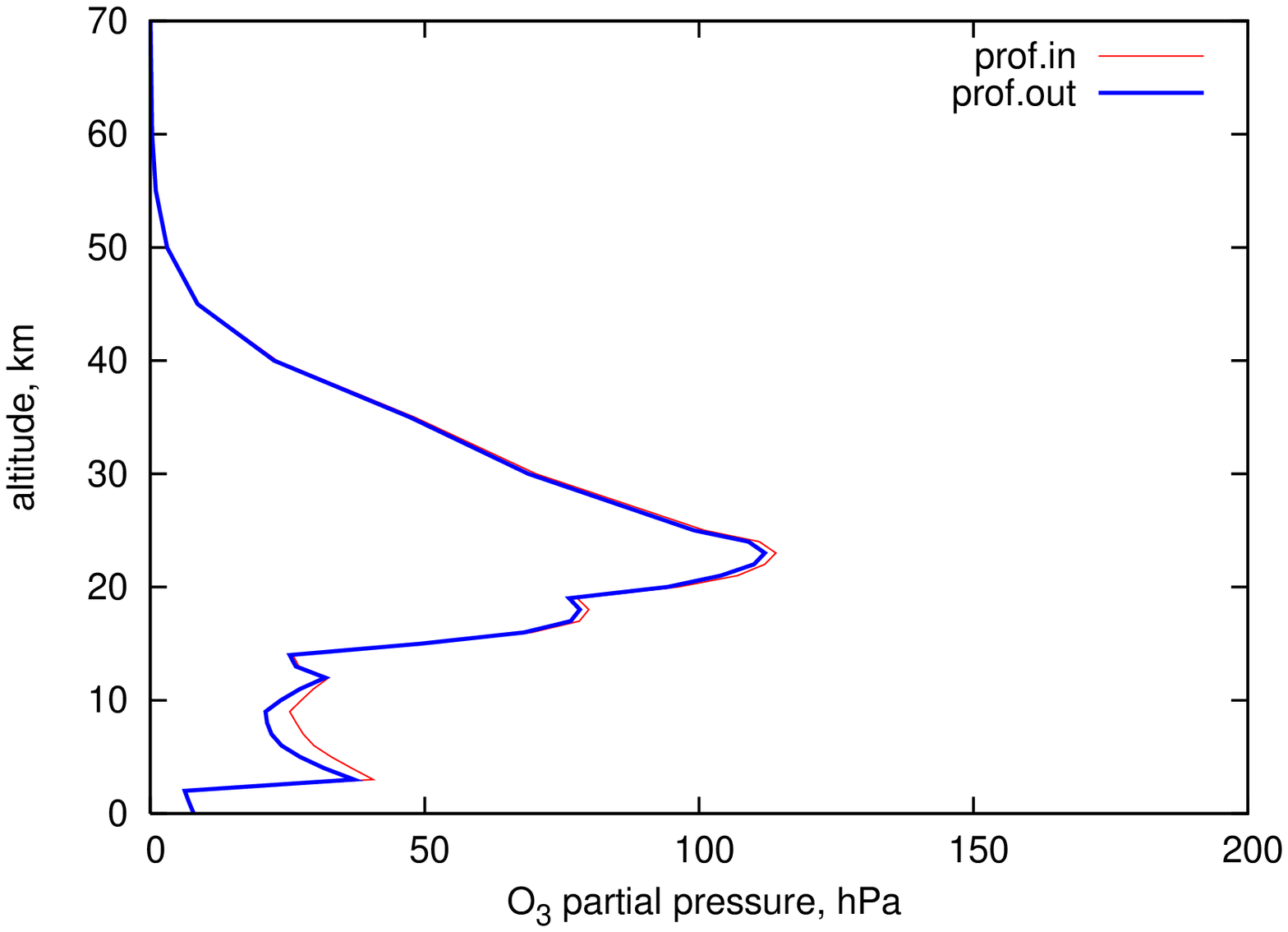}
 \includegraphics[width=8cm]{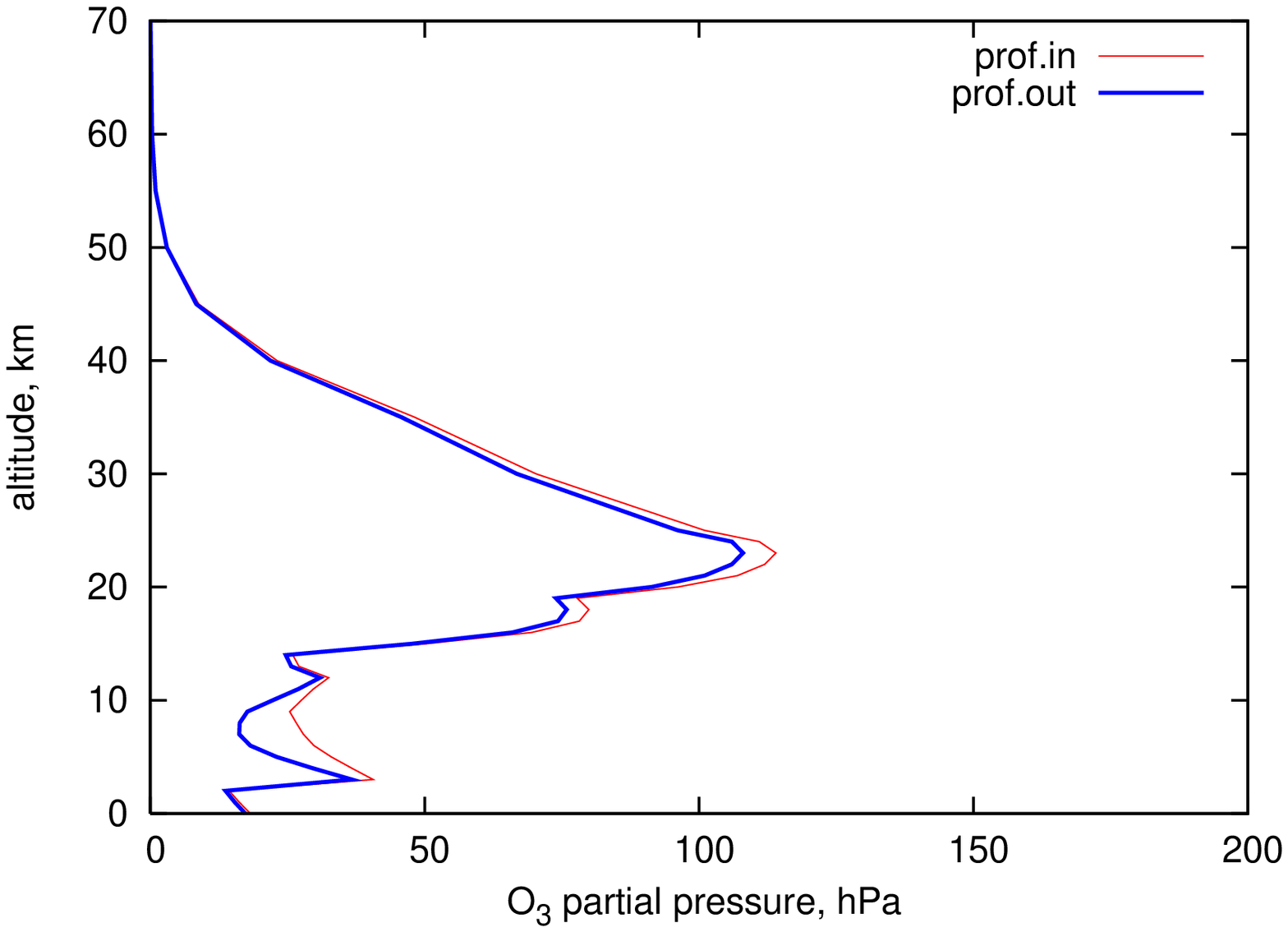}
 \includegraphics[width=8cm]{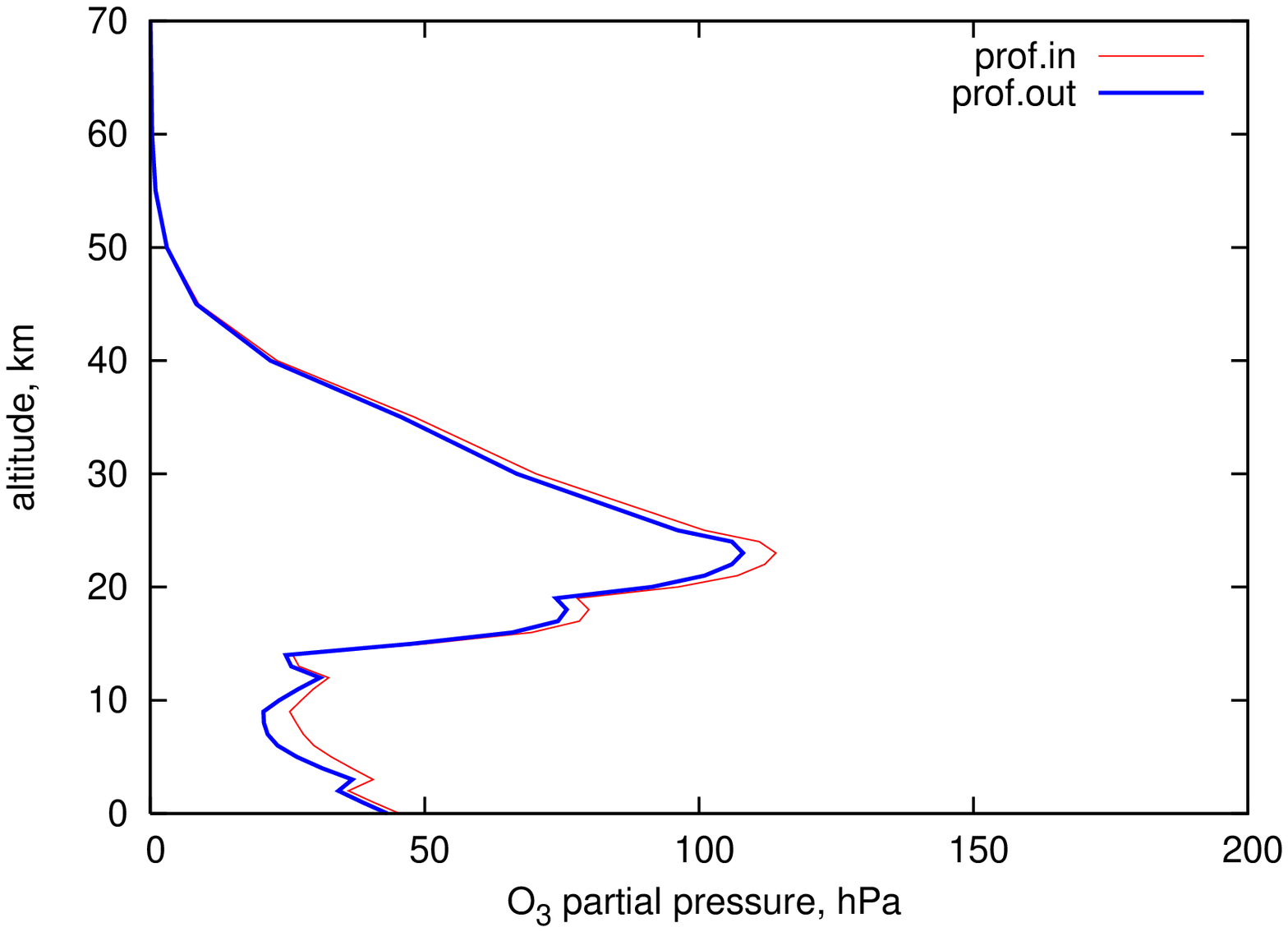}
 \includegraphics[width=8cm]{fig6_3.eps}
      \caption{{\small The retrieved atmospheric ozone profiles for
the 1st o f October 2007 (01.10.07) recorded at 08~h~08~min and
9~h~49~min (upper figures) local time and recorded at
16~h~21~min and 17~h~41~min (lower figures) local time. Please
note that on this day the FTIR total ozone column is rather low:
only 262~DU. Nevertheless, we can see the daily dynamics of
tropospheric ozone: in the morning ozone titration by NO$_x$ is
taking place leading to rather high ozone concentrations later
in the afternoon. Unfortunately, for this day Aura-TES data are
absent and hence the tropospheric part of the input ozone
profile for the MODTRAN modeling process was constructed on the
basis of the TEMIS monthly averaged climatological data.}
              }
         \label{Fig6}
   \end{figure}
%

%%%%%%%%%%%%%%%%%%%%%%%%%%%%%
\section{Conclusion}
%%%%%%%%%%%%%%%%%%%%%%%%%%%%%
%%%%%%%%%%%%%%%%%%%%%%%%%%%%%
We have obtained a long series of  total ozone column estimates
on the base of ground based FTIR observations and MODTRAN
modelling for the years 2005--2007. Our values of the total
ozone columns agree well with OMI satellite remote sensing data.
Differences are in the percentile range. We note some
significant differences under insufficiently clear sky
conditions, which are indicative of the influence of clouds on
FTIR observations.

The work presented here is the first step towards ozone profile
retrievals on a regular basis. For this we need to further
develop our retrieval procedures and we need to perform testing
of our model calculations through ``line-by-line'' radiative
transfer model calculations alike FASCODE. Since we do not have
this code available we need to develop such coding in the near
future ourselves. The procedure of quantitative comparison of
our retrieved profile and other available data must be
developed.\\

{\bf Acknowledgements.} The authors are grateful to the AVDC,
Aura-MLS and Aqua-AIRS website administrations for providing the
necessary satellite remote sensing data. The work of the authors
from MAO NASU was partly supported by the grant of STCU
(2005-2007) and by Space Agency of Ukraine (2007).

%%%%%%%%%%%%%%%%%%%%%%%%%%%%%%%%%%%%%%%%%%%%%%%%%%%%%%%%%%%
%%%%%%%%%%%%%%%%%%%%%%%%%%%%%%%%%%%%%%%%%%%%%%%%%%%%%%%%%%%%


\begin{thebibliography}{99}
%\bibliographystyle{aa}

\bibitem{1}
Bernstein L. S., Berk A., Acharya P. K., Robertson D. C.,
Anderson G. P., Chetwynd J. H., Kimball L. M. Very Narrow Band
Model Calculations of Atmospheric Fluxes and Cooling Rates.
Journal of Atmospheric Sciences. 1996 ,  Vol.53, P.2887--2904.

\bibitem{2}
Bhartia P. K., Wellemeyer C. TOMS-V8 Total O3 Algorithm, OMI
ATBD, Volume II, OMI Ozone Product", P. K. Bhartia, Ed., NASA
GSFC, Greenbelt, MD, OMI-ATBD-02. 2002, P.15--31.

\bibitem{3}
Egevskaya T. B., Vlasov A. M., Bublikov A. V. Infrakrasnyj
Fur'e-spectrometr "Infralum FT-801. Nauka Proizvodstvu. 2001.
n.12, P.38–-41.

\bibitem{4}
Fortuin J., Paul F., Kelder H. An ozone climatology based on
ozonesonde and satellite measurements. Journal of geophysical
research.  1998,  Vol.103,  P.31709--31734.

\bibitem{5}
Kroon M., Brinksma E. J., Labow G., Balis D. OMI-TOMS Total
Ozone Column Validation Status April 2006. RP-OMIE-KNMI-820. May
2006 (Internal KNMI OMI document).

\bibitem{7}
Levelt P. F., Hilsenrath E., Leppelmeier G. W.,  van den Oord
G. H. J., Bhartia P. K., Tamminen J.,  de Haan J. F., Veefkind
J. P. Science Objectives of the Ozone Monitoring Instrument.
IEEE Trans. Geosc. Rem. Sens.  2006,  Vol.44 (5),  P.1199--1208.

\bibitem{8}
OMI README – http://disc.gsfc.nasa.gov/Aura/OMI/.

\bibitem{9}
Perner D., Platt U. Detection of Nitrious Acid in the Atmosphere
by Differential Optical Absorption. J. Geophys. Res.  1979,
Vol.6,  P.917--920.

\bibitem{10}
Rothman L. S., Jasqumart D.,  et al. The HITRAN 2004
molecular spectroscopic database. Journal of Quantitative
Spectroscopy \& Radiative Transfer.  2005,  Vol.96, P.139--204.

\bibitem{11}
Schoeberl M. R.,  Douglass A. R.,  Hilsenrath E.,  Bhartia P.
K., Beer R., Waters J. W., Gunson M. R.,  Froidevaux L.,  Gille
J. C., Barnett J. J., Levelt P. F., and P. DeCola. Overview of
the EOS Aura Mission. IEEE Trans. Geosc. Rem. Sens.  2006,
Vol.44 (5), P.1066--1074.

\bibitem{12}
Shavrina A. V., and Veles A. A. Remote sensing of some
greenhouse gases by Fourier spectrometry in Kyiv. Journal of
Quantitative Spectroscopy \& Radiative Transfer.  2004, Vol.88,
P.345--350.

\bibitem{13}
Shavrina A. V., Pavlenko Ya. V., Veles A., Syniavskyi I., Kroon
M. Ozone columns obtained by ground-based remote sensing in Kiev
for Aura Ozone Measuring Instrument validation. Journal of
geophysical research.  2007,  Vol.113,  P.XXXX.

\bibitem{14}
Veefkind J. P., de Haan J. F., Brinksma E. J., Kroon M., Levelt
P. F., Total ozone from the Ozone Monitoring Instrument (OMI)
using the OMI-DOAS technique. IEEE Trans. Geosc. Rem. Sens.
2006, Vol.44 (5),  P.1239--1244.

\end{thebibliography}
\end{document}